\documentclass[prd,aps,nofootinbib,preprint]{revtex4}
\usepackage{amsmath,amssymb,amsfonts,color,graphicx,graphics,latexsym,placeins,epsfig,multirow}
\usepackage{footmisc}
\usepackage[compatibility=false]{caption}
\usepackage{subcaption}
\usepackage{bbold}
\usepackage{enumitem}
\usepackage{bbold}
\usepackage{hyperref}
\DeclareMathOperator{\sech}{sech}
\usepackage{natbib}
\setcitestyle{square, comma, numbers, sort&compress}
\usepackage{float}
\newcommand{\be}{\begin{equation}}
\newcommand{\ee}{\end{equation}}
\newcommand{\ba}{\begin{eqnarray}}
\newcommand{\ea}{\end{eqnarray}}

\newcommand*{\id}{{\rm\hbox{1\kern-0.15em \vrule width .1pt depth-.2pt}}}

\begin{document}


\title{\Large \bf Memory effects in Kundt wave spacetimes}
\author{Indranil Chakraborty${}^{1}$ and
Sayan Kar ${}^{1,2}$}
\email{indradeb@iitkgp.ac.in, sayan@phy.iitkgp.ac.in}
\affiliation{${}^1$ Centre for Theoretical Studies \\ Indian Institute of Technology Kharagpur, 721 302, India}
\affiliation{${}^{2}$ Department of Physics \\ Indian Institute of Technology Kharagpur, 721 302, India.}

\begin{abstract}
\noindent Memory effects in the exact Kundt wave spacetimes are shown to
arise in the behaviour of geodesics in such spacetimes. The types of Kundt spacetimes we consider here are direct products of the form $H^2\times M(1,1)$ and $S^2\times M(1,1)$. Both geometries have constant scalar curvature. We consider a scenario in which initial velocities of the transverse geodesic coordinates are set to zero (before the arrival of the pulse) in a spacetime with non-vanishing background curvature. We look for changes in the separation between pairs of geodesics caused by the  pulse. Any relative change observed in the position and velocity profiles of geodesics, after the burst, can be solely attributed to the wave (hence, a {\em memory effect}).   
For constant negative curvature, we find there is permanent change in the separation of geodesics after the pulse has departed. Thus, there is displacement memory, though no velocity memory is found.  In the case of constant positive scalar curvature (Pleba\'nski–Hacyan spacetimes), we find both displacement and velocity memory along one direction. In the other  direction, a new kind of memory (which we term as {\em frequency memory effect}) is observed where the {\em separation between the geodesics shows periodic oscillations once the pulse has left}.  We also carry out similar analyses for spacetimes with a non-constant scalar curvature, which may be positive or negative. The results here seem to qualitatively agree with those for constant scalar curvature, thereby suggesting a link between the nature of memory and curvature.
\end{abstract}

\pacs{04.20.-q, 04.20.Jb}
\maketitle

\noindent {\em \textbf {Introduction}}\textemdash The study of memory effects have resurfaced recently since a suggested possibility of detection in advanced detectors like aLIGO, LISA and PTA \citep{Favata:2010},\citep{Lasky:2016}. The frozen-in 'memory' in the  freely falling detectors results in a DC shift in the net relative position (or relative velocity) due to the passage of a gravitational wave. In radiative asymptotically flat spacetimes, this change is related to Bondi-Metzner-Sachs (BMS) \citep{Bondi:1962} transformations relating two inequivalent gravitational vacua. On the theoretical front, a triangular relationship has been conjectured by Strominger \citep{Strominger:2016} relating soft gravitons with BMS symmetries and memories. 

\noindent Historically, the idea of a memory effect 
was first mentioned in a non-resonant detection method put forward by Zel'dovich and Polnarev \citep{Zeldovich:1974} for measuring the radiation emitted from gravitational collapse of stars inside globular clusters. The term 'memory effect' was later coined by Braginsky and Grishchuk \citep{Braginsky:1985} who defined it as the change in the metric perturbation between initial and final times (remote past and distant future). Christodoulou \citep{Christodoulou:1991} generalized the idea of memory in the full nonlinear theory where the stress energy component due to gravitational radiation travels to null infinity. Subsequently, Thorne \citep{Thorne:1991} attributed gravitons, sourced by the outgoing radiation, as being responsible for the nonlinear Christodoulou effect. Bieri, Garfinkle and several others considered  massless fields propagating to null infinity \cite{Tolish:2014} and, consequently,  introduced an electromagnetic (photon) memory effect \citep{Bieri:2013} as well as a neutrino memory effect \citep{Bieri:2015}. Memory effects and their
consequences have been discussed, of late, in cosmology \citep{BieridS:2016,Bieri:2017}, extra dimensional
physics \citep{Wald:2018,Garfinkle:2017,Hollands:2017}, modified gravity \citep{Du:2016,Kilicarslan:2018} and massive gravity \citep{Kilicarslan:2018mass}.

\noindent  
Displacement and velocity memory effects have been addressed by solving geodesic equations for sandwich wave spacetimes \citep{Zhang:2017,Zhang:2017soft,Zhang:2018}. Such spacetimes may contain a localised gravitational wave pulse between two flat Minkowski regions. {\em In this letter, we study memory effects in spacetimes where the two regions separated by the pulse have finite curvature}. Memory effects in radiative spacetimes with non-flat backgrounds have been previously analysed  in \citep{Chu:2019} where the authors have separated the deviation vector arising due to the background curvature and the gravitational wave by solving the geodesic deviation equation in Fermi normal coordinates and isolating the radiative part. Memory in cosmological spacetimes (de Sitter, FRW) \citep{BieridS:2016,Bieri:2017} have been worked out by taking perturbations of the electric and magnetic parts of the Weyl tensor and looking at the behaviour (peel off) of these fields at large spatial distances. Here, we solve the {\em geodesic equations for the spacelike coordinates ($x(u),y(u)$) and look for signatures that a gravitational  wave pulse may imprint on the geodesics, after its departure from the wave region}.

 \noindent We work with non-flat Kundt geometries  \citep{Kundt:1961} as backgrounds. Kundt spacetimes constitute exact radiative solutions of Einstein field equations admitting a nonexpanding, nonshearing and nontwisting null geodesic congruence (NGC). The NGC is not in general covariantly constant. Hence, the wave surfaces are not necessarily Euclidean transverse planes as the null rays are not parallel to each other. Thus, this class of spacetimes are generalisations over {\em pp}  wave spacetimes \citep{Brinkmann:1925, Rosen:1937} where $u$-constant  hypersurfaces are planar. Such geometries have been extensively studied in all algebraically special spacetimes (Petrov classes of type II, III, D, N and O) for vacuum, pure radiation, non-zero cosmological constant and also in higher dimensions (see \citep{Stephani:2003} for a comprehensive review) \citep{Podolski:2001,Ortaggio:2002,Griffiths:2003,Coley:2009,Podolski:2013,Tahamtan:2017}.
 These class of spacetimes  also admit gyratonic solutions \citep{Frolov:2005, Frolov1:2005} which represent localized spinning sources moving at the speed of light. A ring of particles is seen to rotate in presence of a gyraton as they impart angular momentum to the ambient spacetime due to its spin \citep{Frolov:2005}. Such metrics consist of an interior and an exterior region depending upon the presence or absence of gyratonic matter respectively. 
 
 \noindent Throughout this letter, we shall consider the outside region which only contains gravitational radiation. Such waves are known as Kundt gravitational waves which propagate on different spacetime backgrounds \citep{Podolsky:2003,Kadlecova:2009}. The backgrounds that we choose to work with are constant curvature direct-product spacetimes. The two cases studied are $H^2\times M(1,1)$ and $S^2\times M(1,1)$ (Pleba\'nski–Hacyan). We mainly focus on studying geodesic motions in such spacetimes for a sech-squared pulse profile with quadratic dependence on transverse coordinates $x,y$. Such sandwich profiles have been investigated previously by the authors \citep{Chak:2020} while understanding ${\cal B}$-memory in exact plane wave spacetimes. 
 
 \noindent Exact solutions having nonzero twist have been studied with \citep{Bini:2018} or without cosmological constant \citep{Mashhoon:2019} exhibiting nonplanar wavefronts for constant negative Gaussian curvature. These solutions are known as twisted gravitational waves and have influence on the motion of spinning test particles \citep{Bini_spin:2018}. A careful analysis of the geodesics for these two cases reveal significant differences in their nature of memory. For the case with constant positive curvature, a new {\em frequency memory effect} is found which is extensively discussed. To further strengthen the relation between background curvature and the ensuing memory effect, we also study other background spacetimes which have varying scalar curvatures with the same overall sign.

\noindent {\em \textbf {Metric and its matter content}}\textemdash
The general form of a Kundt spacetime is as follows \citep{Griffiths:2009} :
\begin{equation}
ds^2= 2du(dv+W_1dx+W_2dy)+Hdu^2+\frac{1}{P^2}(dx^2+dy^2) \label{eq:kundt_metric}
\end{equation}
$P\equiv  P(u,x,y), H \equiv H(u,v,x,y), W_i \equiv W_i(u,v,x,y), $ $\forall \hspace{2mm} i  \hspace{2mm}\epsilon \hspace{2mm} \{x,y\}$.

\noindent The NGC is given by a vector field $\mathbf{k}$, where $\mathbf{k}=\partial_v$. $\mathbf{k}$ is also orthogonal to tangent vector fields to the spatial wave surfaces. We do not consider gyratonic matter present and hence $W_i$ is always set to zero. We will also assume $H$ is independent of $v$. Thus, the metric line element used is:  
\begin{equation}
ds^2= 2dudv+Hdu^2+\frac{1}{P^2}(dx^2+dy^2) \label{eq:kundt_wave_metric}
\end{equation}

\noindent The Ricci scalar and the only non zero Einstein tensor for the Kundt wave metric given in Eq.(\ref{eq:kundt_wave_metric}) turn out as following:
\begin{align}
   {\cal R} =2\Delta(\ln P)  \label{eq:ricci_scalar}\\
  G^u\,_u=-\Delta(\ln P)     \label{eq:einstein_tensor}
\end{align}
\noindent Here, $\Delta=P^2(\partial_{xx}+\partial_{yy})$ \citep{Tahamtan:2017}. For spacetimes having non vanishing cosmological constant and pure radiation (electromagnetic fields) \citep{Kadlecova:2009}, the Einstein field equations for the $G^u\,_u$ component (see Eq.(\ref{eq:einstein_tensor}))become
\begin{equation}
\Delta(\ln P)=\Lambda+\rho \label{eq:EFE}
\end{equation}
\noindent Since the L. H. S. of Eq.(\ref{eq:EFE}) involves derivatives of only transverse $x,y$ coordinates and its R. H. S. is a constant, there is no loss of generality in setting $P$ independent of $u$. The canonical choice for $P$ given by Eq.(\ref{eq:EFE}) generates wave surfaces of constant curvature \citep{Griffiths:2009}. In our letter,  we consider both the positive ($S^2$) and negative ($H^2$) constant curvature cases separately.\footnote{The zero case corresponds to the $pp$ wave metric.} Hence, the metric (\ref{eq:kundt_wave_metric}) can be thought as gravitational waves propagating over a background geometry. The background spacetimes that we work with are direct-product constant curvature spacetimes \textendash $H^2\times M(1,1)$ and $S^2\times M(1,1)$ (Pleba\'nski–Hacyan spacetimes \citep{Ortaggio:2002}).\footnote{$M(1,1)$ denotes minkowski spacetime of 1+1 dimension} Eq.(\ref{eq:einstein_tensor}) shows the presence of a non-vanishing energy momentum tensor following from Einstein field equations. This matter content is nonzero and finite at every $u$-constant hypersurface. {\em The presence of matter allows these hypersurfaces to have finite curvature (as is evident from Eq.(\ref{eq:ricci_scalar})) and hence they are not the usual Euclidean transverse planes}.\footnote{This is unlike pp-wave spacetime where the hypersurfaces are exact Euclidean transverse planes having zero spatial curvature.} Here, we choose $H=\sech^2(u)(x^2-y^2)$ and $P\equiv P(x,y)$ acts as the square root of the inverse conformal factor determining the curvature of spatial 2-surfaces. Initially, we choose $P$ so that the geometry is $S^2$ or $H^2$. Later, we will relax this constant curvature criterion and examine the behaviour of geodesics in variable curvature metrics having a fixed sign for the curvature.

\vspace*{1cm}

\noindent {\em \textbf {Kundt Waves in} $\mathbf{H^2\times M(1,1)}$}\textemdash The functional form of $P(x,y)=y$. The invariant Ricci scalar is $-2$ (negative). The metric takes the form:

\begin{equation}  
  ds^2=2dudv+\sech^2(u)(x^2-y^2)du^2+\frac{1}{y^2}(dx^2+dy^2) \label{eq:metric_neg_curv}
\end{equation}

\noindent For the metric given in Eq.(\ref{eq:metric_neg_curv}) the geodesic equations for $x,y$ are:

\begin{gather}
\frac{d^2x}{du^2}-\frac{2}{y}\bigg(\frac{dx}{du}\bigg)\bigg(\frac{dy}{du}\bigg)-\sech^2(u)xy^2=0  \label{eq:x_neg_curv}\\
\frac{d^2y}{du^2}-\frac{1}{y}\bigg(\frac{dy}{du}\bigg)^2+\frac{1}{y}\bigg(\frac{dx}{du}\bigg)^2 +\sech^2(u)y^3=0 \label{eq:y_neg_curv}
\end{gather}
\noindent Note that in the asymptotic regions (i.e. $u \rightarrow \pm \infty$),
constant $x$ and $y$ solve both the equations. This happens even though the
spacetime has a negative scalar curvature.

\begin{figure}[H]
	\centering
	\begin{subfigure}[t]{0.45\textwidth}
		\centering
		\includegraphics[width=\textwidth]{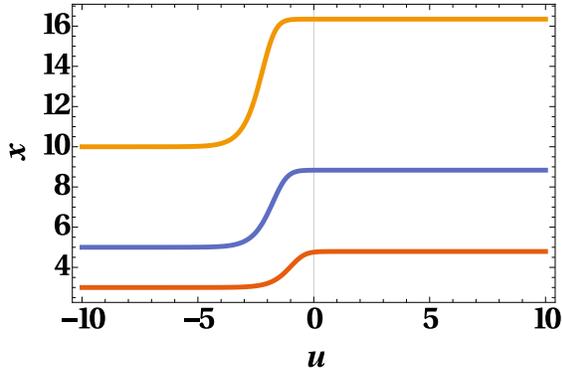}
		 \caption{\centering{\small Initial positions of $x$ are 3(orange), 5(blue)and 10(yellow) respectively.}}
		\label{fig:x_neg_curv}
\end{subfigure}\hspace{1cm}
	\begin{subfigure}[t]{0.45\textwidth}
		\centering
		\includegraphics[width=\textwidth]{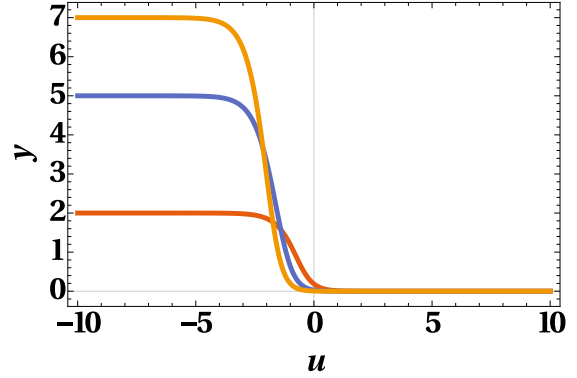}
		 \caption{\centering{\small Initial positions of $y$ are 2(orange), 5(blue)and 7(yellow) respectively.}}
		\label{fig:y_neg_curv}
	\end{subfigure}
	\caption{Displacement memory effect in Kundt waves with spatial 2-surfaces having negative curvature.}
	\label{fig:Disp_neg_curv}
\end{figure}

\noindent The geodesic equations cannot be solved analytically and hence they are solved numerically using \textit{Mathematica 10}. For this class of metrics, $u$ acts as an affine parameter. We analyse geodesics and plot the solution of geodesic equations along the transverse spatial directions. The plots are given in Figs. \ref{fig:x_neg_curv} and \ref{fig:y_neg_curv}.

\noindent The plots in Figs.(\ref{fig:x_neg_curv}) and (\ref{fig:y_neg_curv}) demonstrate that the geodesics have permanent separation after encountering the gravitational wave pulse. The change in position before and after is dependent on the initial position of the pulse. We observe that the separation increases along $x$  while it decreases along $y$.

\begin{figure}[H]
	\centering
	\begin{subfigure}[t]{0.45\textwidth}
		\centering
		\includegraphics[width=\textwidth]{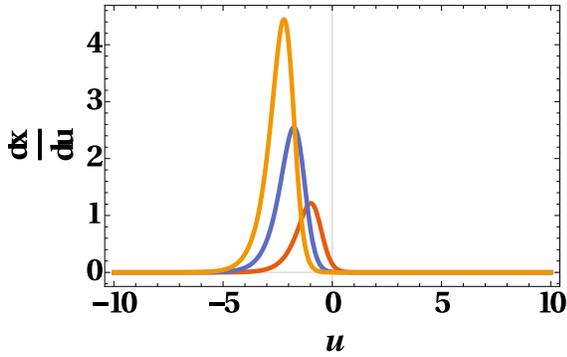}
		 \caption{\centering{$x$ direction}}
		\label{fig:x_vel_neg_curv}
\end{subfigure}\hspace{1.5cm}
	\begin{subfigure}[t]{0.45\textwidth}
		\centering
		\includegraphics[width=\textwidth]{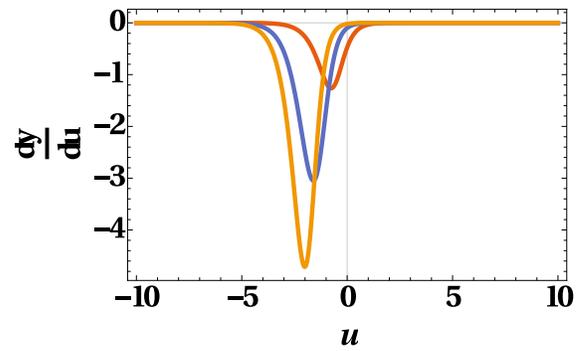}
		 \caption{\centering{$y$ direction}}
		\label{fig:y_vel_neg_curv}
	\end{subfigure}
	\caption{Velocity memory effect for Kundt waves with spatial 2-surfaces having negative curvature.}
	\label{fig:Vel_neg_curv}
\end{figure}

\noindent We now look at the analysis of velocity of the geodesics from the plots shown in Figs. (\ref{fig:x_vel_neg_curv}) and (\ref{fig:y_vel_neg_curv}).

\noindent Fig.(\ref{fig:Vel_neg_curv}) shows that there is no velocity memory effect in these spacetimes. There is a transient peak corresponding to the presence of the pulse. After the departure of the pulse, the velocity vanishes. Thus along both directions ({\em i.e.} x \& y), there is no change in the velocity of the timelike\footnote{In these class of metrics, the geodesic equations for $x(u),y(u)$ are same for both timelike and null.} geodesics at early and late times.

\noindent Hence, from this metric we can clearly observe displacement memory effect as proposed by Zel'dovich and Polnarev and others \citep{Zeldovich:1974},\citep{Braginsky:1985}.

\noindent {\em \textbf {Kundt Waves in} $\mathbf{S^2\times M(1,1)}$}\textemdash The conformal factor becomes $\sech^2(y)$
({\em i.e.} $P=\cosh{y}$). The metric takes up the form:

\begin{equation}  
  ds^2=2dudv+\sech^2(u)(x^2-y^2)du^2+\sech^2(y)(dx^2+dy^2) \label{eq:metric_posv_curv}
\end{equation}

\noindent The geodesic equations for the transverse spatial coordinates are:

 \begin{gather}
  \frac{d^2 x}{du^2}-2\tanh(y)\bigg(\frac{dx}{du}\bigg)\bigg(\frac{dy}{du}\bigg) -x\frac{\sech^2(u)}{\sech^2(y)}=0 \label{eq:x_posv_curv}\\
  \frac{d^2 y}{du^2}-\tanh(y)\bigg(\frac{dy}{du}\bigg)^2+\tanh(y)\bigg(\frac{dx}{du}\bigg)^2+y\frac{\sech^2(u)}{\sech^2(y)}=0 \label{eq:y_posv_curv}
 \end{gather}

\begin{figure}[H]
	\centering
	\begin{subfigure}[t]{0.45\textwidth}
		\centering
		\includegraphics[width=\textwidth]{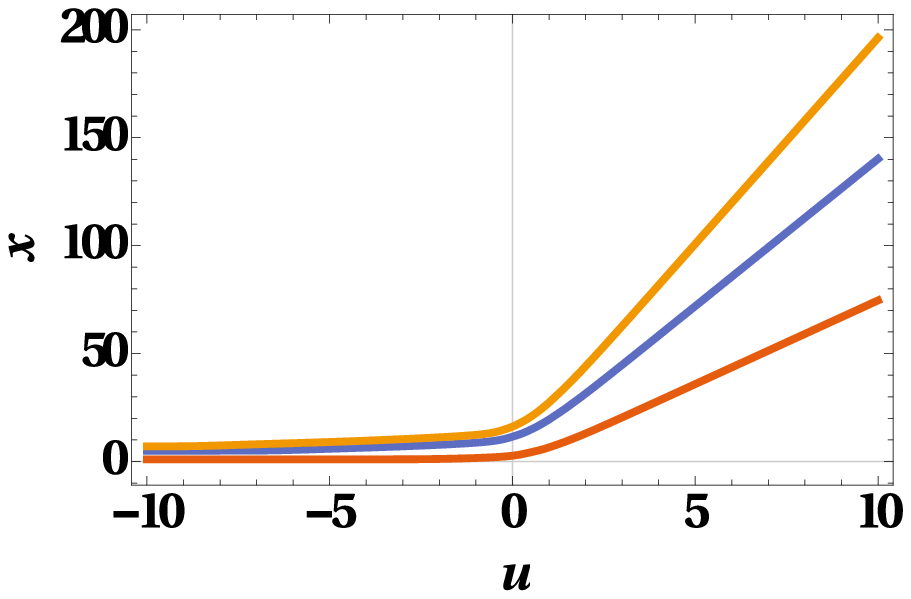}
		 \caption{\centering{\large $x$ direction}}
		\label{fig:x_pos_curv}
\end{subfigure}\hspace{1.5cm}
	\begin{subfigure}[t]{0.45\textwidth}
		\centering
		\includegraphics[width=\textwidth]{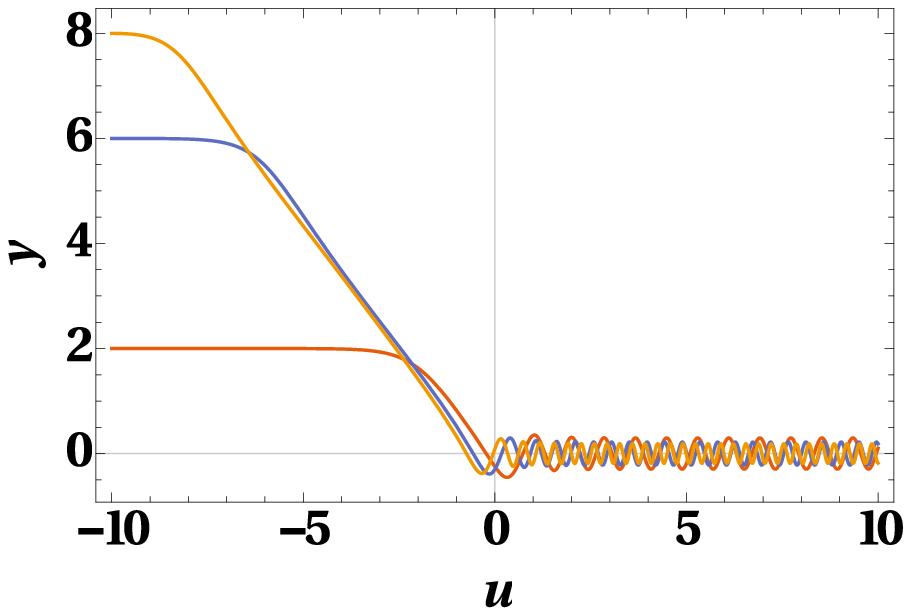}
		 \caption{\centering{\large $y$ direction}}
		\label{fig:y_pos_curv}
	\end{subfigure}
	\caption{Displacement memory effect in Kundt waves for constant positive spatial 2-surfaces with initial positions as $(x,y)=\{(1,2),(5,6),(7,8)\}$ for the three geodesics respectively denoted in orange, blue and yellow respectively.}
	\label{fig:Disp_pos_curv}
\end{figure}

\noindent We were unable to find any analytical solution for Eqs.(\ref{eq:x_posv_curv}) and (\ref{eq:y_posv_curv}). Hence, they are solved numerically using {\em Mathematica 10}. The plots are given in Figs. (\ref{fig:x_pos_curv}) and (\ref{fig:y_pos_curv}).

\noindent In Fig.(\ref{fig:Disp_pos_curv}), we observe different types of memory along different axes. For the $x$ direction given in Fig.(\ref{fig:x_pos_curv}), there is increase in separation after the passage of the pulse. But along the $y$ direction, we find in Fig.(\ref{fig:y_pos_curv}) that the geodesics tend to oscillate about zero value. Thus, the separation between them also oscillates. This is a new phenomenon which has not yet been discussed in the previous literature on memory. We call it as the {\em frequency memory effect}. The reason for calling it as a memory effect is because there is a characteristic frequency with which the separation oscillates after the passage of the pulse. We will study this effect extensively later. Next, we look into the velocities of the trajectories.

\begin{figure}[H]
	\centering
	\begin{subfigure}[t]{0.45\textwidth}
		\centering
		\includegraphics[width=\textwidth]{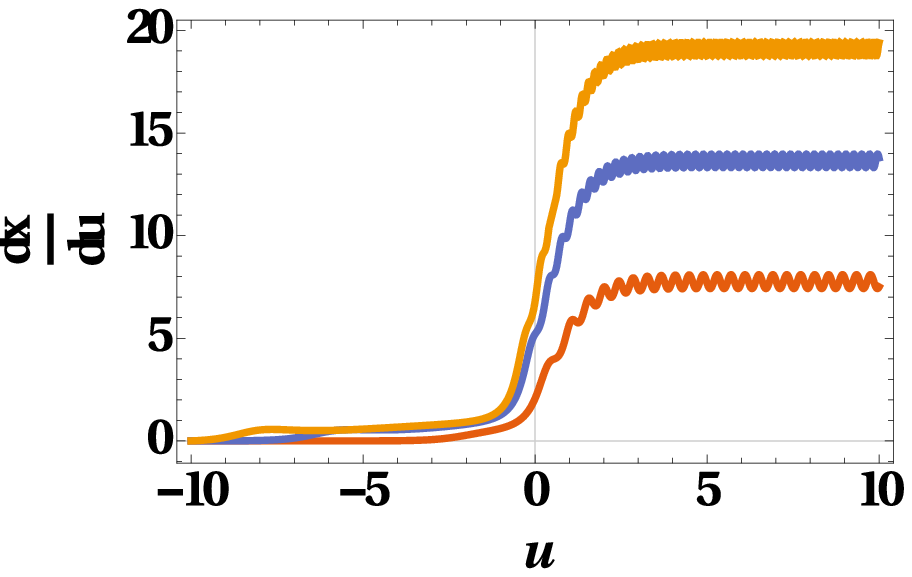}
		 \caption{\centering{\large $x$ direction}}
		\label{fig:x_vel_pos_curv}
\end{subfigure}\hspace{1.0cm}
	\begin{subfigure}[t]{0.45\textwidth}
		\centering
		\includegraphics[width=\textwidth]{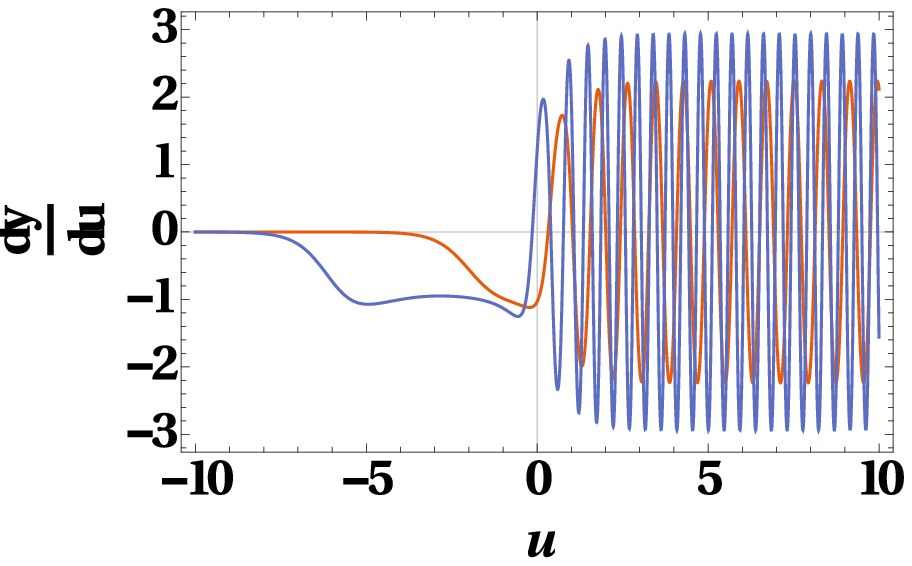}
		 \caption{\centering{\large $y$ direction}}
		\label{fig:y_vel_pos_curv}
	\end{subfigure}
	\caption{Velocity memory effect for Kundt waves with spatial 2-surfaces having positive curvature. In the $x$ direction, the three colours denote the velocities of the geodesics having the same initial velocities as given in Fig.(\ref{fig:x_pos_curv}). In the $y$ direction we have only plotted two geodesics corresponding to the first two values of initial positions, for better clarity.}
	\label{fig:Vel_pos_curv}
\end{figure}

\noindent The plots in Fig. (\ref{fig:Vel_pos_curv}) shows there is velocity memory present. In the $x$ direction, the final velocity settles to a fixed value. While in y direction, there is periodical fluctuation owing to the frequency memory effect. Moreover, we find difference in amplitudes because the geodesics which seem to oscillate more frequently have higher magnitudes of velocity. \footnote{While taking  time derivative over the phases of a wave, the frequency component come out as a factor.}  

\noindent {\em \textbf{Frequency memory effect}}\textemdash  The oscillatory behaviour of geodesics after the passage of pulse in Fig.(\ref{fig:y_pos_curv}) calls for further attention. Hence, we examine the plots with three different initial positions and look at the nature of separation of nearby geodesics.

\begin{figure}[H]
	\centering
	\begin{subfigure}[t]{0.4\textwidth}
		\centering
		\includegraphics[width=\textwidth]{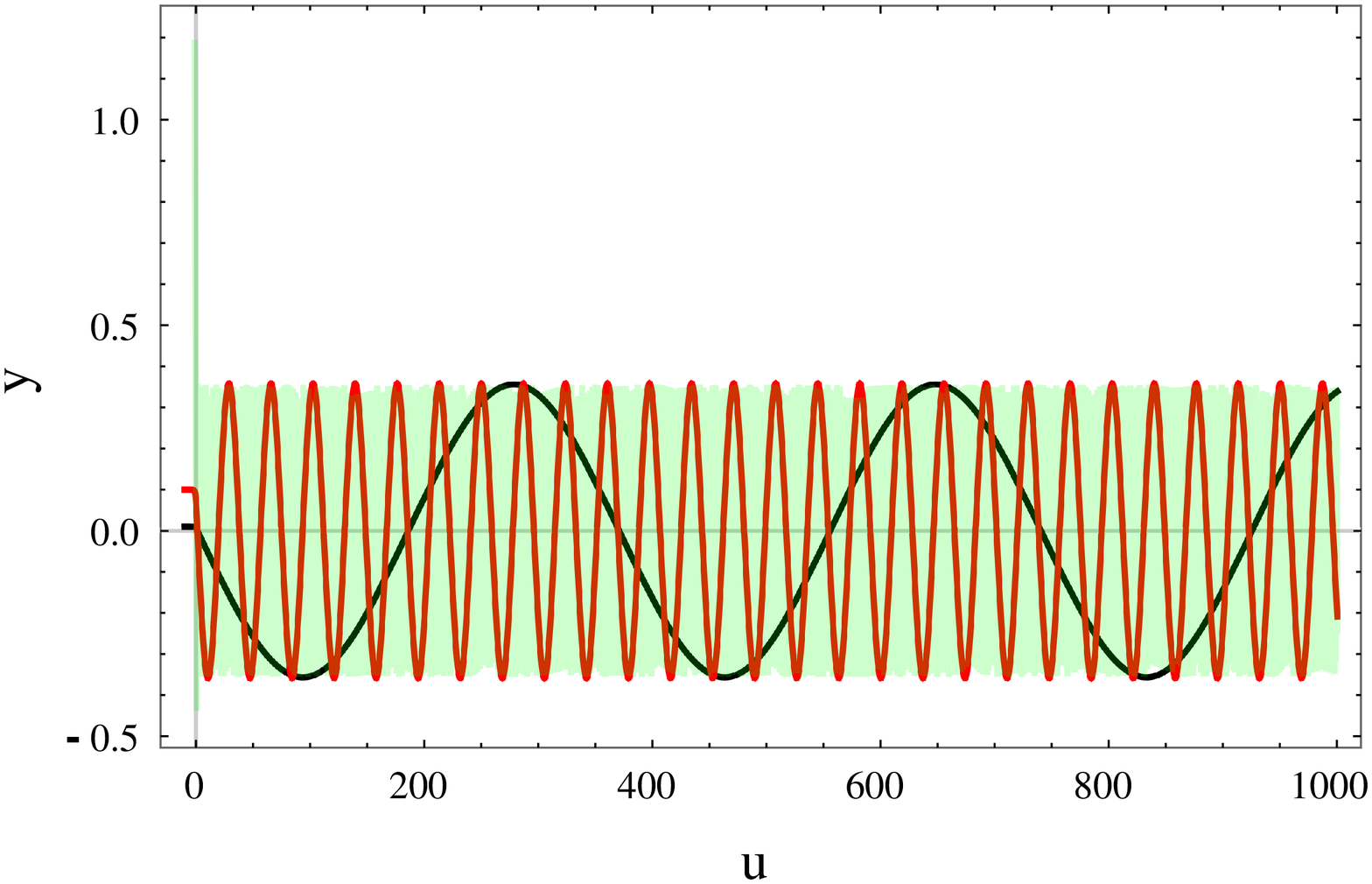}
		 \caption{\centering{Oscillatory behaviour analyzed for three different initial separations.}}
		\label{fig:oscillation}
\end{subfigure}\hspace{1cm}
	\begin{subfigure}[t]{0.4\textwidth}
		\centering
		\includegraphics[width=\textwidth]{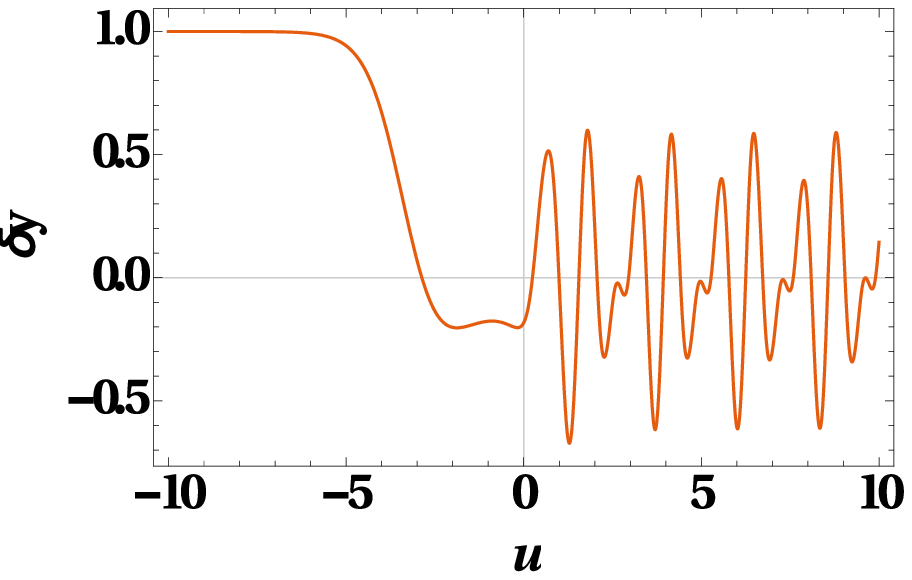}
		 \caption{Separation along y direction with initial coordinates $[\{3,4\},\{2,3\}]$, resulting in formation of beats.}
		\label{fig:beats_memory}
	\end{subfigure}
	\caption{Frequency memory effect.}
	\label{fig:frequency_memory_effect}
\end{figure}

\noindent Fig.(\ref{fig:oscillation}) reveals the behaviour of geodesics post departure of the pulse for three different initial configurations. In the following table we give the details of the frequency (wavelength) of oscillation for three separate initial positions.

\begin{center}

\begin{table}[!h]
    
\begin{tabular}{|c|c|c|c|c|c|c|} 
  \hline
$x$ & $y$ & 1st Trough & 2nd Trough  & Frequency ($u^{-1}$) & $\lambda$ (in units of $u$)  & Colour\\
\hline
   2.0      & 3.0       & 5.98       & 7.13   & 0.869565    & 1.15 &  Light green\\
  
  0.1  &  0.1  &  28.82  & 65.67 & 0.027137 & 36.85 & Red\\
  
  0.01 & 0.01 & 278.5 & 647.9 & 0.002707 & 369.4 & Black\\
  \hline
  \end{tabular}
  \label{Tab:oscillation}
  \caption{ Frequency and wavelength of oscillation corresponding to different initial separations.}
  \end{table}  
\end{center}
\vspace*{-1cm}
\noindent The distance between two successive troughs/crests is denoted as the wavelength (in units of $u$). The frequency is taken as the inverse of the wavelength. From the above Table it is clear that higher value of $y$-coordinate initially gives a high frequency oscillation (shown in light green) while smaller values yield smaller frequencies (shown in black in Fig.(\ref{fig:oscillation})). 

\noindent There seems to be an occurrence of beats when we consider the evolution of geodesic deviation as shown in Fig.(\ref{fig:beats_memory}). The variation in the pattern is due to the difference in $y$-coordinate values of each geodesic. Thus, one can attribute a particular frequency to a single geodesic solution. Hence, it can be said that for positive spatial curvature there is a particular frequency associated
with the geodesic caused entirely by the passage of a gravitational wave pulse. 
Similar to displacement and velocity memory, we may name this oscillatory 
feature as a {\em frequency memory effect}.

\noindent {\em \textbf{Pulse nature}}\textemdash In our entire analysis we have used sech-squared as the nature of our pulse profile. However, the nature of memory is quite independent of the nature of the chosen profile. In fact, for any such sandwich spacetimes (we have checked for the Gaussian pulse profile) one can observe permanent displacement as well as oscillatory behaviour for constant negative curvature and positive curvature respectively.

 \noindent {\em \textbf{Variable curvature}}
\textemdash The geodesic motion studied so far were in constant curvature direct-product spacetimes. The results obtained show that there are distinct classes of memory for different background curvatures. Thus, it is essential to analyse geodesics in varying curvature spacetimes to understand the connection between memory and curvature. We choose certain functional forms for $P$ that yields varying scalar curvature with an overall
positive or negative sign for the metric (\ref{eq:kundt_wave_metric}).   

\noindent {\em Negative curvature}\textemdash In order to achieve variable negative curvature we take $P=\sech y$ (scalar curvature, $R=-2\sech^4 y$), for which the two-dimensional
$xy$ space is that of the catenoid. The geodesic equations are as follows:

\begin{gather}
  \frac{d^2 x}{du^2}+2\tanh(y)\bigg(\frac{dx}{du}\bigg)\bigg(\frac{dy}{du}\bigg) -x\frac{\sech^2(u)}{\cosh^2(y)}=0 \label{eq:x_var_neg_curv}\\
  \frac{d^2 y}{du^2}+\tanh(y)\bigg(\frac{dy}{du}\bigg)^2-\tanh(y)\bigg(\frac{dx}{du}\bigg)^2+y\frac{\sech^2(u)}{\cosh^2(y)}=0 \label{eq:y_var_neg_curv}
 \end{gather}

\begin{figure}[H]
	\centering
	\begin{subfigure}[t]{0.37\textwidth}
		\centering
		\includegraphics[width=\textwidth]{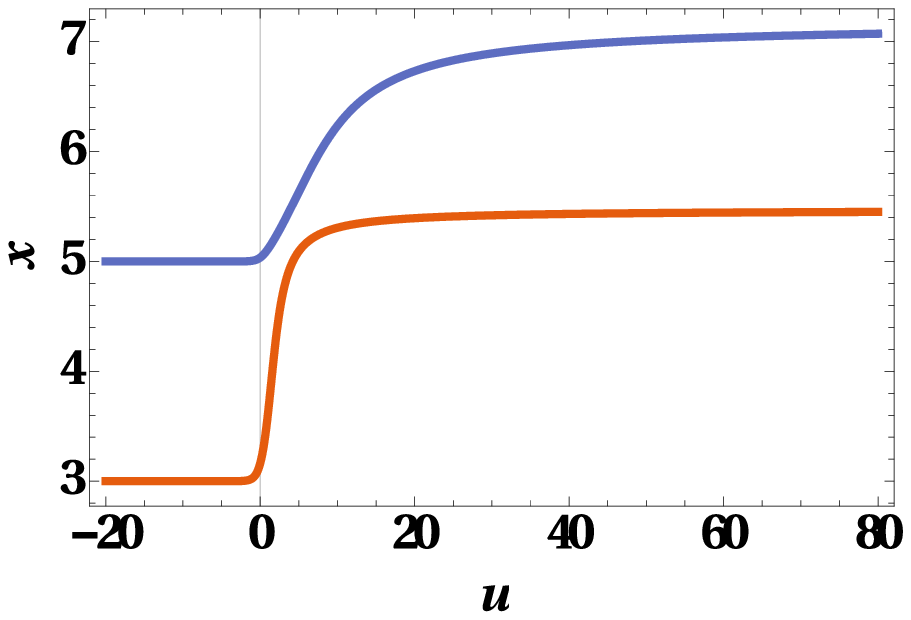}
		 \caption{\centering{$x$ direction}}
		\label{fig:x_var_neg_curv}
\end{subfigure}\hspace{2cm}
	\begin{subfigure}[t]{0.38\textwidth}
		\centering
		\includegraphics[width=\textwidth]{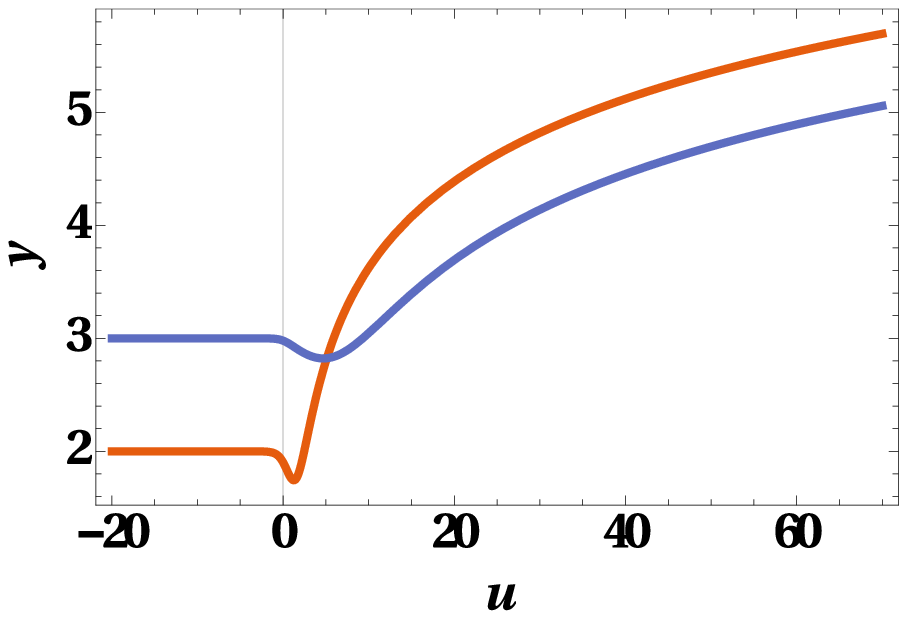}
		 \caption{\centering{$y$ direction}}
		\label{fig:y_var_neg_curv}
	\end{subfigure}
	\caption{Displacement memory effect for Kundt waves with spatial 2-surfaces having variable negative curvature with initial positions of the geodesics as $\{(3,2),(5,3)\}$ for two geodesics denoted by orange and blue colours respectively.}
	\label{fig:Var_neg_curv}
\end{figure}

\noindent In Fig.(\ref{fig:x_var_neg_curv}) we see that initially parallel geodesics, after the passage of the pulse, undergoes a change in separation along $x$, which is almost constant. As seen in  Fig.(\ref{fig:y_var_neg_curv}),  for the $y$ coordinate
there is an intersection but finally there is again a constant separation with a non zero slope. This nature is similar to the plots in Fig.(\ref{fig:Disp_neg_curv}) indicating that negative curvature solutions produce only displacement memory . 

\noindent {\em Positive curvature}\textemdash In order to have variable positive curvature on the two dimensional $xy$ space, we take $P=\sec{y}$ (scalar curvature, $R= 2\sec^4{y} >0$). The geodesic equations become:

\begin{gather}
  \frac{d^2 x}{du^2}-2\tanh(y)\bigg(\frac{dx}{du}\bigg)\bigg(\frac{dy}{du}\bigg) -x\frac{\sech^2(u)}{\cos^2(y)}=0 \label{eq:x_var_posv_curv}\\
  \frac{d^2 y}{du^2}-\tanh(y)\bigg(\frac{dy}{du}\bigg)^2+\tanh(y)\bigg(\frac{dx}{du}\bigg)^2+y\frac{\sech^2(u)}{\cos^2(y)}=0 \label{eq:y_var_posv_curv}
 \end{gather}

\begin{figure}[H]
	\centering
	\begin{subfigure}[t]{0.4\textwidth}
		\centering
		\includegraphics[width=\textwidth]{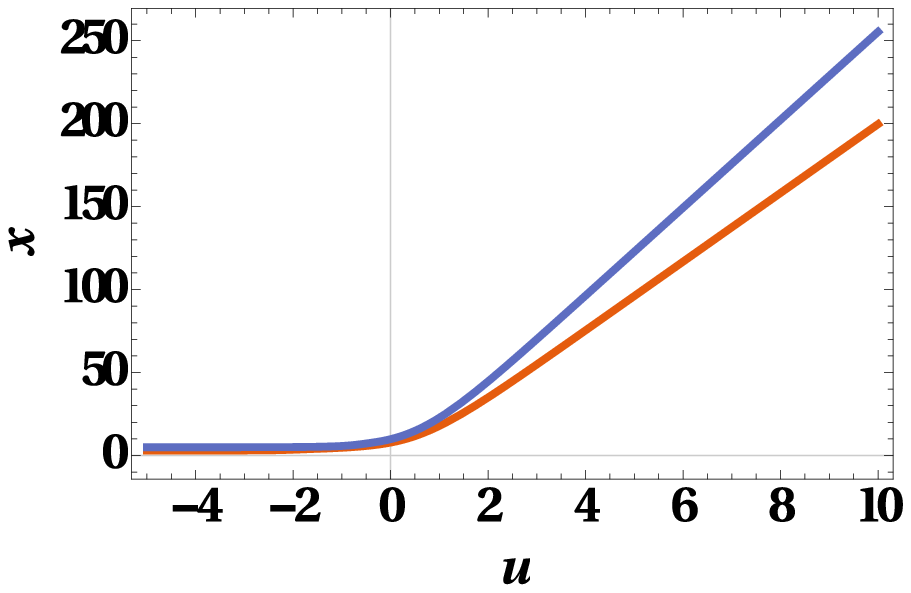}
		 \caption{\centering{$x$ direction}}
		\label{fig:x_var_posv_curv}
\end{subfigure}\hspace{2cm}
	\begin{subfigure}[t]{0.4\textwidth}
		\centering
		\includegraphics[width=\textwidth]{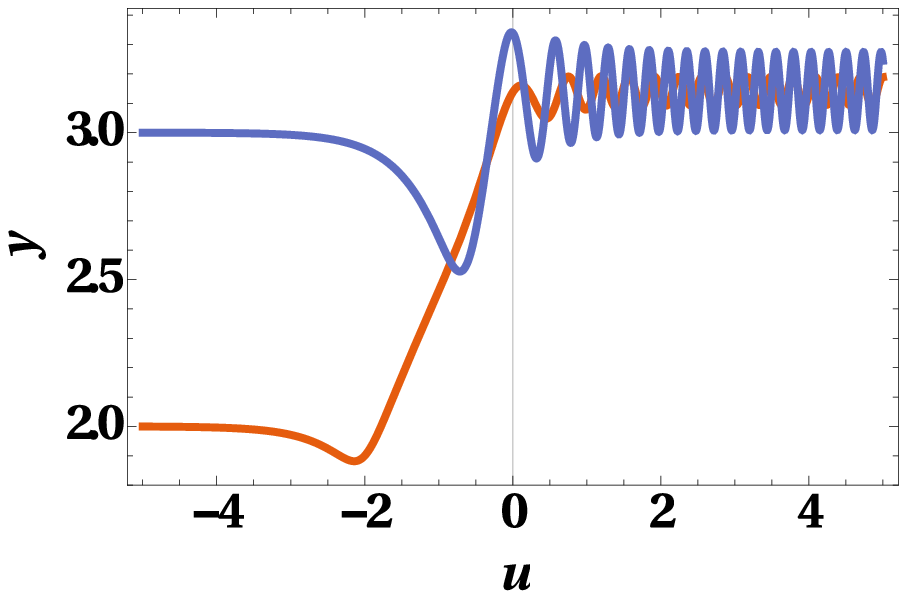}
		 \caption{\centering{$y$ direction}}
		\label{fig:y_var_posv_curv}
	\end{subfigure}
	\caption{Displacement memory effect for Kundt waves with spatial 2-surfaces having variable positive curvature with initial positions of the geodesics as $\{(3,2),(5,3)\}$ for two geodesics shown in orange and blue colours,  respectively.}
	\label{fig:Var_posv_curv}
\end{figure}

\noindent Fig.(\ref{fig:x_var_posv_curv}) shows monotonic increase in displacement memory. In Fig.(\ref{fig:y_var_posv_curv}) we find the same oscillatory behaviour as found earlier in Fig.(\ref{fig:y_pos_curv}). This once again corresponds to the fact that positive curvature solutions produce a frequency memory effect through
an oscillatory behaviour after the pulse has departed.

\noindent {\em \textbf{Conclusions}}\textemdash Memory effects in Kundt wave spacetimes have been worked out in this letter. The method we adopt is to analyse the evolution of timelike geodesics before and after they encounter a localised gravitational wave source. The wave propagates on background (direct-product) spacetimes $S^2\times M(1,1)$ and $H^2\times M(1,1)$. We consider a metric in which the entire curvature comes from the two dimensional induced metric of the transverse wave surfaces. Using Einstein equations, we try to motivate the reason for spatial curvature in the wave surfaces. We find that matter is present on every $u$-constant hypersurface, which results
in it being curved. 

\noindent The geodesic solutions are obtained numerically and plotted. We find displacement memory in $H^2\times M(1,1)$ along both $x,y$ directions. In $S^2\times M(1,1)$, we find displacement and velocity memory along $x$ direction while a new {\em frequency memory effect} appears in the $y$ direction. The term {\em memory} is attributed to this new phenomenon because there is a characteristic periodic oscillation of the geodesics caused due to the gravitational wave pulse. Frequency memory is explicitly studied for three different initial configurations. We observe that higher initial value of the coordinate yields higher frequencies (low wavelengths) and vice versa. Three different cases are tabulated (Table I) for better quantitative understanding.

\noindent Our entire analysis is independent of the choice of sandwich pulse profiles. For example if we assume a general Gaussian pulse (checked, but not shown here) we recover the same nature of memory as for the sech-squared pulse discussed above. We also explore variable curvature scenarios which have same sign. It is noted that similar qualitative features are retained in the latter case where curvature
is not constant.

\noindent Thus, through studies on the memory effect one may understand the
behaviour of test particle motion in
various background spacetimes of the Kundt class in a better way. Studying ${\cal B}$-memory \citep{Loughlin:2019,Chak:2020,Srijit:2019} of geodesic congruences in Kundt wave spacetimes will perhaps reveal another type of memory. It will be of interest
to know if a similar frequency memory effect arises for $\cal B$-memory too.  

\noindent {\em \textbf{Acknowledgements}}\textemdash I. C. acknowledges the University Grants Commission (UGC) of the Government of India for  providing  financial  assistance  through  senior research fellowship (SRF) with reference ID: 523711.

\bibliographystyle{apsrev4-2}
\bibliography{mybibliography_2}

\begin{thebibliography}{44}%
\makeatletter
\providecommand \@ifxundefined [1]{%
 \@ifx{#1\undefined}
}%
\providecommand \@ifnum [1]{%
 \ifnum #1\expandafter \@firstoftwo
 \else \expandafter \@secondoftwo
 \fi
}%
\providecommand \@ifx [1]{%
 \ifx #1\expandafter \@firstoftwo
 \else \expandafter \@secondoftwo
 \fi
}%
\providecommand \natexlab [1]{#1}%
\providecommand \enquote  [1]{``#1''}%
\providecommand \bibnamefont  [1]{#1}%
\providecommand \bibfnamefont [1]{#1}%
\providecommand \citenamefont [1]{#1}%
\providecommand \href@noop [0]{\@secondoftwo}%
\providecommand \href [0]{\begingroup \@sanitize@url \@href}%
\providecommand \@href[1]{\@@startlink{#1}\@@href}%
\providecommand \@@href[1]{\endgroup#1\@@endlink}%
\providecommand \@sanitize@url [0]{\catcode `\\12\catcode `\$12\catcode
  `\&12\catcode `\#12\catcode `\^12\catcode `\_12\catcode `\%12\relax}%
\providecommand \@@startlink[1]{}%
\providecommand \@@endlink[0]{}%
\providecommand \url  [0]{\begingroup\@sanitize@url \@url }%
\providecommand \@url [1]{\endgroup\@href {#1}{\urlprefix }}%
\providecommand \urlprefix  [0]{URL }%
\providecommand \Eprint [0]{\href }%
\providecommand \doibase [0]{https://doi.org/}%
\providecommand \selectlanguage [0]{\@gobble}%
\providecommand \bibinfo  [0]{\@secondoftwo}%
\providecommand \bibfield  [0]{\@secondoftwo}%
\providecommand \translation [1]{[#1]}%
\providecommand \BibitemOpen [0]{}%
\providecommand \bibitemStop [0]{}%
\providecommand \bibitemNoStop [0]{.\EOS\space}%
\providecommand \EOS [0]{\spacefactor3000\relax}%
\providecommand \BibitemShut  [1]{\csname bibitem#1\endcsname}%
\let\auto@bib@innerbib\@empty
\bibitem [{\citenamefont {Favata}(2010)}]{Favata:2010}%
  \BibitemOpen
  \bibfield  {author} {\bibinfo {author} {\bibfnamefont {M.}~\bibnamefont
  {Favata}},\ }\href {https://doi.org/10.1088/0264-9381/27/8/084036} {\bibfield
   {journal} {\bibinfo  {journal} {Class. Qtm. Grav.}\ }\textbf {\bibinfo
  {volume} {27}},\ \bibinfo {pages} {084036} (\bibinfo {year}
  {2010})}\BibitemShut {NoStop}%
\bibitem [{\citenamefont {Lasky}\ \emph {et~al.}(2016)\citenamefont {Lasky},
  \citenamefont {Thrane}, \citenamefont {Levin}, \citenamefont {Blackman},\
  and\ \citenamefont {Chen}}]{Lasky:2016}%
  \BibitemOpen
  \bibfield  {author} {\bibinfo {author} {\bibfnamefont {P.~D.}\ \bibnamefont
  {Lasky}}, \bibinfo {author} {\bibfnamefont {E.}~\bibnamefont {Thrane}},
  \bibinfo {author} {\bibfnamefont {Y.}~\bibnamefont {Levin}}, \bibinfo
  {author} {\bibfnamefont {J.}~\bibnamefont {Blackman}},\ and\ \bibinfo
  {author} {\bibfnamefont {Y.}~\bibnamefont {Chen}},\ }\href
  {https://doi.org/10.1103/PhysRevLett.117.061102} {\bibfield  {journal}
  {\bibinfo  {journal} {Phys. Rev. Lett.}\ }\textbf {\bibinfo {volume} {117}},\
  \bibinfo {pages} {061102} (\bibinfo {year} {2016})}\BibitemShut {NoStop}%
\bibitem [{\citenamefont {Bondi}\ \emph {et~al.}(1962)\citenamefont {Bondi},
  \citenamefont {van~der Burg},\ and\ \citenamefont {Metzner}}]{Bondi:1962}%
  \BibitemOpen
  \bibfield  {author} {\bibinfo {author} {\bibfnamefont {H.}~\bibnamefont
  {Bondi}}, \bibinfo {author} {\bibfnamefont {M.~G.~J.}\ \bibnamefont {van~der
  Burg}},\ and\ \bibinfo {author} {\bibfnamefont {A.~W.~K.}\ \bibnamefont
  {Metzner}},\ }\href {https://doi.org/10.1098/rspa.1962.0161} {\bibfield
  {journal} {\bibinfo  {journal} {Proc. Roy. Soc. Lond. A}\ }\textbf {\bibinfo
  {volume} {269}},\ \bibinfo {pages} {21} (\bibinfo {year} {1962})}\BibitemShut
  {NoStop}%
\bibitem [{\citenamefont {Strominger}\ and\ \citenamefont
  {Zhiboedov}()}]{Strominger:2016}%
  \BibitemOpen
  \bibfield  {author} {\bibinfo {author} {\bibfnamefont {A.}~\bibnamefont
  {Strominger}}\ and\ \bibinfo {author} {\bibfnamefont {A.}~\bibnamefont
  {Zhiboedov}},\ }\href {https://doi.org/10.1007/JHEP01(2016)086} {\bibfield
  {journal} {\bibinfo  {journal} {J. High Energy Phys.}\ }\textbf {\bibinfo
  {volume} {01}}\bibinfo  {number} { (2016)},\ \bibinfo {pages}
  {86}}\BibitemShut {NoStop}%
\bibitem [{\citenamefont {{Zel'dovich}}\ and\ \citenamefont
  {{Polnarev}}(1974)}]{Zeldovich:1974}%
  \BibitemOpen
\bibfield  {number} {  }\bibfield  {author} {\bibinfo {author} {\bibfnamefont
  {Y.~B.}\ \bibnamefont {{Zel'dovich}}}\ and\ \bibinfo {author} {\bibfnamefont
  {A.~G.}\ \bibnamefont {{Polnarev}}},\ }\href@noop {} {\bibfield  {journal}
  {\bibinfo  {journal} {Sov. Astron}\ }\textbf {\bibinfo {volume} {18}},\
  \bibinfo {pages} {17} (\bibinfo {year} {1974})}\BibitemShut {NoStop}%
\bibitem [{\citenamefont {Braginsky}\ and\ \citenamefont
  {Grishchuk}(1985)}]{Braginsky:1985}%
  \BibitemOpen
  \bibfield  {author} {\bibinfo {author} {\bibfnamefont {V.~B.}\ \bibnamefont
  {Braginsky}}\ and\ \bibinfo {author} {\bibfnamefont {L.~P.}\ \bibnamefont
  {Grishchuk}},\ }\href@noop {} {\bibfield  {journal} {\bibinfo  {journal}
  {Sov. Phys. JETP}\ }\textbf {\bibinfo {volume} {62}},\ \bibinfo {pages} {427}
  (\bibinfo {year} {1985})}\BibitemShut {NoStop}%
\bibitem [{\citenamefont {Christodoulou}(1991)}]{Christodoulou:1991}%
  \BibitemOpen
  \bibfield  {author} {\bibinfo {author} {\bibfnamefont {D.}~\bibnamefont
  {Christodoulou}},\ }\href {https://doi.org/10.1103/PhysRevLett.67.1486}
  {\bibfield  {journal} {\bibinfo  {journal} {Phys. Rev. Lett.}\ }\textbf
  {\bibinfo {volume} {67}},\ \bibinfo {pages} {1486} (\bibinfo {year}
  {1991})}\BibitemShut {NoStop}%
\bibitem [{\citenamefont {Thorne}(1992)}]{Thorne:1991}%
  \BibitemOpen
  \bibfield  {author} {\bibinfo {author} {\bibfnamefont {K.~S.}\ \bibnamefont
  {Thorne}},\ }\href {https://doi.org/10.1103/PhysRevD.45.520} {\bibfield
  {journal} {\bibinfo  {journal} {Phys. Rev. D}\ }\textbf {\bibinfo {volume}
  {45}},\ \bibinfo {pages} {520} (\bibinfo {year} {1992})}\BibitemShut
  {NoStop}%
\bibitem [{\citenamefont {Tolish}\ and\ \citenamefont
  {Wald}(2014)}]{Tolish:2014}%
  \BibitemOpen
  \bibfield  {author} {\bibinfo {author} {\bibfnamefont {A.}~\bibnamefont
  {Tolish}}\ and\ \bibinfo {author} {\bibfnamefont {R.~M.}\ \bibnamefont
  {Wald}},\ }\href {https://doi.org/10.1103/PhysRevD.89.064008} {\bibfield
  {journal} {\bibinfo  {journal} {Phys. Rev. D}\ }\textbf {\bibinfo {volume}
  {89}},\ \bibinfo {pages} {064008} (\bibinfo {year} {2014})}\BibitemShut
  {NoStop}%
\bibitem [{\citenamefont {Bieri}\ and\ \citenamefont
  {Garfinkle}(2013)}]{Bieri:2013}%
  \BibitemOpen
  \bibfield  {author} {\bibinfo {author} {\bibfnamefont {L.}~\bibnamefont
  {Bieri}}\ and\ \bibinfo {author} {\bibfnamefont {D.}~\bibnamefont
  {Garfinkle}},\ }\href {https://doi.org/10.1088/0264-9381/30/19/195009}
  {\bibfield  {journal} {\bibinfo  {journal} {Class. Qtm. Grav.}\ }\textbf
  {\bibinfo {volume} {30}},\ \bibinfo {pages} {195009} (\bibinfo {year}
  {2013})}\BibitemShut {NoStop}%
\bibitem [{\citenamefont {Bieri}\ and\ \citenamefont
  {Garfinkle}(2015)}]{Bieri:2015}%
  \BibitemOpen
  \bibfield  {author} {\bibinfo {author} {\bibfnamefont {L.}~\bibnamefont
  {Bieri}}\ and\ \bibinfo {author} {\bibfnamefont {D.}~\bibnamefont
  {Garfinkle}},\ }\href {https://doi.org/10.1007/s00023-014-0329-1} {\bibfield
  {journal} {\bibinfo  {journal} {Ann. Henri Poincar{\'e}}\ }\textbf {\bibinfo
  {volume} {16}},\ \bibinfo {pages} {801} (\bibinfo {year} {2015})}\BibitemShut
  {NoStop}%
\bibitem [{\citenamefont {Bieri}\ \emph {et~al.}(2016)\citenamefont {Bieri},
  \citenamefont {Garfinkle},\ and\ \citenamefont {Yau}}]{BieridS:2016}%
  \BibitemOpen
  \bibfield  {author} {\bibinfo {author} {\bibfnamefont {L.}~\bibnamefont
  {Bieri}}, \bibinfo {author} {\bibfnamefont {D.}~\bibnamefont {Garfinkle}},\
  and\ \bibinfo {author} {\bibfnamefont {S.-T.}\ \bibnamefont {Yau}},\ }\href
  {https://doi.org/10.1103/PhysRevD.94.064040} {\bibfield  {journal} {\bibinfo
  {journal} {Phys. Rev. D}\ }\textbf {\bibinfo {volume} {94}},\ \bibinfo
  {pages} {064040} (\bibinfo {year} {2016})}\BibitemShut {NoStop}%
\bibitem [{\citenamefont {Bieri}\ \emph {et~al.}(2017)\citenamefont {Bieri},
  \citenamefont {Garfinkle},\ and\ \citenamefont {Yunes}}]{Bieri:2017}%
  \BibitemOpen
  \bibfield  {author} {\bibinfo {author} {\bibfnamefont {L.}~\bibnamefont
  {Bieri}}, \bibinfo {author} {\bibfnamefont {D.}~\bibnamefont {Garfinkle}},\
  and\ \bibinfo {author} {\bibfnamefont {N.}~\bibnamefont {Yunes}},\ }\href
  {https://doi.org/10.1088/1361-6382/aa8b52} {\bibfield  {journal} {\bibinfo
  {journal} {Class. Qtm. Grav.}\ }\textbf {\bibinfo {volume} {34}},\ \bibinfo
  {pages} {215002} (\bibinfo {year} {2017})}\BibitemShut {NoStop}%
\bibitem [{\citenamefont {Satishchandran}\ and\ \citenamefont
  {Wald}(2018)}]{Wald:2018}%
  \BibitemOpen
  \bibfield  {author} {\bibinfo {author} {\bibfnamefont {G.}~\bibnamefont
  {Satishchandran}}\ and\ \bibinfo {author} {\bibfnamefont {R.~M.}\
  \bibnamefont {Wald}},\ }\href {https://doi.org/10.1103/PhysRevD.97.024036}
  {\bibfield  {journal} {\bibinfo  {journal} {Phys. Rev. D}\ }\textbf {\bibinfo
  {volume} {97}},\ \bibinfo {pages} {024036} (\bibinfo {year}
  {2018})}\BibitemShut {NoStop}%
\bibitem [{\citenamefont {Garfinkle}\ \emph {et~al.}(2017)\citenamefont
  {Garfinkle}, \citenamefont {Hollands}, \citenamefont {Ishibashi},
  \citenamefont {Tolish},\ and\ \citenamefont {Wald}}]{Garfinkle:2017}%
  \BibitemOpen
  \bibfield  {author} {\bibinfo {author} {\bibfnamefont {D.}~\bibnamefont
  {Garfinkle}}, \bibinfo {author} {\bibfnamefont {S.}~\bibnamefont {Hollands}},
  \bibinfo {author} {\bibfnamefont {A.}~\bibnamefont {Ishibashi}}, \bibinfo
  {author} {\bibfnamefont {A.}~\bibnamefont {Tolish}},\ and\ \bibinfo {author}
  {\bibfnamefont {R.~M.}\ \bibnamefont {Wald}},\ }\href
  {https://doi.org/10.1088/1361-6382/aa777b} {\bibfield  {journal} {\bibinfo
  {journal} {Classical and Quantum Gravity}\ }\textbf {\bibinfo {volume}
  {34}},\ \bibinfo {pages} {145015} (\bibinfo {year} {2017})}\BibitemShut
  {NoStop}%
\bibitem [{\citenamefont {Hollands}\ \emph {et~al.}(2017)\citenamefont
  {Hollands}, \citenamefont {Ishibashi},\ and\ \citenamefont
  {Wald}}]{Hollands:2017}%
  \BibitemOpen
  \bibfield  {author} {\bibinfo {author} {\bibfnamefont {S.}~\bibnamefont
  {Hollands}}, \bibinfo {author} {\bibfnamefont {A.}~\bibnamefont
  {Ishibashi}},\ and\ \bibinfo {author} {\bibfnamefont {R.~M.}\ \bibnamefont
  {Wald}},\ }\href {https://doi.org/10.1088/1361-6382/aa777a} {\bibfield
  {journal} {\bibinfo  {journal} {Classical and Quantum Gravity}\ }\textbf
  {\bibinfo {volume} {34}},\ \bibinfo {pages} {155005} (\bibinfo {year}
  {2017})}\BibitemShut {NoStop}%
\bibitem [{\citenamefont {Du}\ and\ \citenamefont {Nishizawa}(2016)}]{Du:2016}%
  \BibitemOpen
  \bibfield  {author} {\bibinfo {author} {\bibfnamefont {S.~M.}\ \bibnamefont
  {Du}}\ and\ \bibinfo {author} {\bibfnamefont {A.}~\bibnamefont {Nishizawa}},\
  }\href {https://doi.org/10.1103/PhysRevD.94.104063} {\bibfield  {journal}
  {\bibinfo  {journal} {Phys. Rev. D}\ }\textbf {\bibinfo {volume} {94}},\
  \bibinfo {pages} {104063} (\bibinfo {year} {2016})}\BibitemShut {NoStop}%
\bibitem [{\citenamefont {Kilicarslan}(2019)}]{Kilicarslan:2018}%
  \BibitemOpen
  \bibfield  {author} {\bibinfo {author} {\bibfnamefont {E.}~\bibnamefont
  {Kilicarslan}},\ }\href {https://doi.org/10.3906/fiz-1811-2} {\bibfield
  {journal} {\bibinfo  {journal} {Turk. J. Phys.}\ }\textbf {\bibinfo {volume}
  {43}},\ \bibinfo {pages} {126} (\bibinfo {year} {2019})}\BibitemShut
  {NoStop}%
\bibitem [{\citenamefont {Kilicarslan}\ and\ \citenamefont
  {Tekin}(2019)}]{Kilicarslan:2018mass}%
  \BibitemOpen
  \bibfield  {author} {\bibinfo {author} {\bibfnamefont {E.}~\bibnamefont
  {Kilicarslan}}\ and\ \bibinfo {author} {\bibfnamefont {B.}~\bibnamefont
  {Tekin}},\ }\href {https://doi.org/10.1140/epjc/s10052-019-6636-4} {\bibfield
   {journal} {\bibinfo  {journal} {Eur. Phys. J.}\ }\textbf {\bibinfo {volume}
  {C79}},\ \bibinfo {pages} {114} (\bibinfo {year} {2019})}\BibitemShut
  {NoStop}%
\bibitem [{\citenamefont {Zhang}\ \emph
  {et~al.}(2017{\natexlab{a}})\citenamefont {Zhang}, \citenamefont {Duval},
  \citenamefont {Gibbons},\ and\ \citenamefont {Horvathy}}]{Zhang:2017}%
  \BibitemOpen
  \bibfield  {author} {\bibinfo {author} {\bibfnamefont {P.-M.}\ \bibnamefont
  {Zhang}}, \bibinfo {author} {\bibfnamefont {C.}~\bibnamefont {Duval}},
  \bibinfo {author} {\bibfnamefont {G.~W.}\ \bibnamefont {Gibbons}},\ and\
  \bibinfo {author} {\bibfnamefont {P.~A.}\ \bibnamefont {Horvathy}},\ }\href
  {https://doi.org/10.1016/j.physletb.2017.07.050} {\bibfield  {journal}
  {\bibinfo  {journal} {Phys. Lett. B}\ }\textbf {\bibinfo {volume} {772}},\
  \bibinfo {pages} {743} (\bibinfo {year} {2017}{\natexlab{a}})}\BibitemShut
  {NoStop}%
\bibitem [{\citenamefont {Zhang}\ \emph
  {et~al.}(2017{\natexlab{b}})\citenamefont {Zhang}, \citenamefont {Duval},
  \citenamefont {Gibbons},\ and\ \citenamefont {Horvathy}}]{Zhang:2017soft}%
  \BibitemOpen
  \bibfield  {author} {\bibinfo {author} {\bibfnamefont {P.-M.}\ \bibnamefont
  {Zhang}}, \bibinfo {author} {\bibfnamefont {C.}~\bibnamefont {Duval}},
  \bibinfo {author} {\bibfnamefont {G.~W.}\ \bibnamefont {Gibbons}},\ and\
  \bibinfo {author} {\bibfnamefont {P.~A.}\ \bibnamefont {Horvathy}},\ }\href
  {https://doi.org/10.1103/PhysRevD.96.064013} {\bibfield  {journal} {\bibinfo
  {journal} {Phys. Rev. D}\ }\textbf {\bibinfo {volume} {96}},\ \bibinfo
  {pages} {064013} (\bibinfo {year} {2017}{\natexlab{b}})}\BibitemShut
  {NoStop}%
\bibitem [{\citenamefont {Zhang}\ \emph {et~al.}(2018)\citenamefont {Zhang},
  \citenamefont {Duval},\ and\ \citenamefont {Horvathy}}]{Zhang:2018}%
  \BibitemOpen
  \bibfield  {author} {\bibinfo {author} {\bibfnamefont {P.-M.}\ \bibnamefont
  {Zhang}}, \bibinfo {author} {\bibfnamefont {C.}~\bibnamefont {Duval}},\ and\
  \bibinfo {author} {\bibfnamefont {P.~A.}\ \bibnamefont {Horvathy}},\ }\href
  {https://doi.org/10.1088/1361-6382/aaa987} {\bibfield  {journal} {\bibinfo
  {journal} {Class. Qtm. Grav.}\ }\textbf {\bibinfo {volume} {35}},\ \bibinfo
  {pages} {065011} (\bibinfo {year} {2018})}\BibitemShut {NoStop}%
\bibitem [{\citenamefont {Chu}\ and\ \citenamefont {Koyama}(2019)}]{Chu:2019}%
  \BibitemOpen
  \bibfield  {author} {\bibinfo {author} {\bibfnamefont {C.-S.}\ \bibnamefont
  {Chu}}\ and\ \bibinfo {author} {\bibfnamefont {Y.}~\bibnamefont {Koyama}},\
  }\href {https://doi.org/10.1103/PhysRevD.100.104034} {\bibfield  {journal}
  {\bibinfo  {journal} {Phys. Rev. D}\ }\textbf {\bibinfo {volume} {100}},\
  \bibinfo {pages} {104034} (\bibinfo {year} {2019})}\BibitemShut {NoStop}%
\bibitem [{\citenamefont {Kundt}(1961)}]{Kundt:1961}%
  \BibitemOpen
  \bibfield  {author} {\bibinfo {author} {\bibfnamefont {W.}~\bibnamefont
  {Kundt}},\ }\href {https://doi.org/10.1007/BF01328918} {\bibfield  {journal}
  {\bibinfo  {journal} {Z. Phys.}\ }\textbf {\bibinfo {volume} {163}},\
  \bibinfo {pages} {77} (\bibinfo {year} {1961})}\BibitemShut {NoStop}%
\bibitem [{\citenamefont {Brinkmann}(1925)}]{Brinkmann:1925}%
  \BibitemOpen
  \bibfield  {author} {\bibinfo {author} {\bibfnamefont {H.~W.}\ \bibnamefont
  {Brinkmann}},\ }\href {https://doi.org/10.1007/BF01208647} {\bibfield
  {journal} {\bibinfo  {journal} {Math. Ann.}\ }\textbf {\bibinfo {volume}
  {94}},\ \bibinfo {pages} {119} (\bibinfo {year} {1925})}\BibitemShut
  {NoStop}%
\bibitem [{\citenamefont {Rosen}(1937)}]{Rosen:1937}%
  \BibitemOpen
  \bibfield  {author} {\bibinfo {author} {\bibfnamefont {N.}~\bibnamefont
  {Rosen}},\ }\href@noop {} {\bibfield  {journal} {\bibinfo  {journal} {Phys.
  Z. Sowjetunion}\ }\textbf {\bibinfo {volume} {12}},\ \bibinfo {pages} {366}
  (\bibinfo {year} {1937})}\BibitemShut {NoStop}%
\bibitem [{\citenamefont {Stephani}\ \emph {et~al.}(2003)\citenamefont
  {Stephani}, \citenamefont {Kramer}, \citenamefont {MacCallum}, \citenamefont
  {Hoenselaers},\ and\ \citenamefont {Herlt}}]{Stephani:2003}%
  \BibitemOpen
  \bibfield  {author} {\bibinfo {author} {\bibfnamefont {H.}~\bibnamefont
  {Stephani}}, \bibinfo {author} {\bibfnamefont {D.}~\bibnamefont {Kramer}},
  \bibinfo {author} {\bibfnamefont {M.~A.~H.}\ \bibnamefont {MacCallum}},
  \bibinfo {author} {\bibfnamefont {C.}~\bibnamefont {Hoenselaers}},\ and\
  \bibinfo {author} {\bibfnamefont {E.}~\bibnamefont {Herlt}},\ }\href
  {https://doi.org/10.1017/CBO9780511535185} {\emph {\bibinfo {title} {{Exact
  solutions of Einstein's field equations}}}}\ (\bibinfo  {publisher}
  {Cambridge Univ. Press},\ \bibinfo {address} {Cambridge, England},\ \bibinfo
  {year} {2003})\BibitemShut {NoStop}%
\bibitem [{\citenamefont {Podolsk{\'{y}}}\ and\ \citenamefont
  {Ortaggio}(2001)}]{Podolski:2001}%
  \BibitemOpen
  \bibfield  {author} {\bibinfo {author} {\bibfnamefont {J.}~\bibnamefont
  {Podolsk{\'{y}}}}\ and\ \bibinfo {author} {\bibfnamefont {M.}~\bibnamefont
  {Ortaggio}},\ }\href {https://doi.org/10.1088/0264-9381/18/14/307} {\bibfield
   {journal} {\bibinfo  {journal} {Classical and Quantum Gravity}\ }\textbf
  {\bibinfo {volume} {18}},\ \bibinfo {pages} {2689} (\bibinfo {year}
  {2001})}\BibitemShut {NoStop}%
\bibitem [{\citenamefont {{M. Ortaggio and J.
  Podolsk{\'{y}}}}(2002)}]{Ortaggio:2002}%
  \BibitemOpen
  \bibfield  {author} {\bibinfo {author} {\bibnamefont {{M. Ortaggio and J.
  Podolsk{\'{y}}}}},\ }\href {https://doi.org/10.1088/0264-9381/19/20/313}
  {\bibfield  {journal} {\bibinfo  {journal} {Classical and Quantum Gravity}\
  }\textbf {\bibinfo {volume} {19}},\ \bibinfo {pages} {5221} (\bibinfo {year}
  {2002})}\BibitemShut {NoStop}%
\bibitem [{\citenamefont {Griffiths}\ \emph {et~al.}(2003)\citenamefont
  {Griffiths}, \citenamefont {Docherty},\ and\ \citenamefont
  {Podolsk{\'{y}}}}]{Griffiths:2003}%
  \BibitemOpen
  \bibfield  {author} {\bibinfo {author} {\bibfnamefont {J.~B.}\ \bibnamefont
  {Griffiths}}, \bibinfo {author} {\bibfnamefont {P.}~\bibnamefont
  {Docherty}},\ and\ \bibinfo {author} {\bibfnamefont {J.}~\bibnamefont
  {Podolsk{\'{y}}}},\ }\href {https://doi.org/10.1088/0264-9381/21/1/014}
  {\bibfield  {journal} {\bibinfo  {journal} {Class. Qtm. Grav.}\ }\textbf
  {\bibinfo {volume} {21}},\ \bibinfo {pages} {207} (\bibinfo {year}
  {2003})}\BibitemShut {NoStop}%
\bibitem [{\citenamefont {Coley}\ \emph {et~al.}(2009)\citenamefont {Coley},
  \citenamefont {Hervik}, \citenamefont {Papadopoulos},\ and\ \citenamefont
  {Pelavas}}]{Coley:2009}%
  \BibitemOpen
  \bibfield  {author} {\bibinfo {author} {\bibfnamefont {A.}~\bibnamefont
  {Coley}}, \bibinfo {author} {\bibfnamefont {S.}~\bibnamefont {Hervik}},
  \bibinfo {author} {\bibfnamefont {G.}~\bibnamefont {Papadopoulos}},\ and\
  \bibinfo {author} {\bibfnamefont {N.}~\bibnamefont {Pelavas}},\ }\href
  {https://doi.org/10.1088/0264-9381/26/10/105016} {\bibfield  {journal}
  {\bibinfo  {journal} {Class. Qtm. Grav.}\ }\textbf {\bibinfo {volume} {26}},\
  \bibinfo {pages} {105016} (\bibinfo {year} {2009})}\BibitemShut {NoStop}%
\bibitem [{\citenamefont {Podolsk{\'{y}}}\ and\ \citenamefont
  {{\v{S}}varc}(2013)}]{Podolski:2013}%
  \BibitemOpen
  \bibfield  {author} {\bibinfo {author} {\bibfnamefont {J.}~\bibnamefont
  {Podolsk{\'{y}}}}\ and\ \bibinfo {author} {\bibfnamefont {R.}~\bibnamefont
  {{\v{S}}varc}},\ }\href {https://doi.org/10.1088/0264-9381/30/20/205016}
  {\bibfield  {journal} {\bibinfo  {journal} {Class. Qtm. Grav.}\ }\textbf
  {\bibinfo {volume} {30}},\ \bibinfo {pages} {205016} (\bibinfo {year}
  {2013})}\BibitemShut {NoStop}%
\bibitem [{\citenamefont {Tahamtan}\ and\ \citenamefont
  {Sv{\'i}tek}(2017)}]{Tahamtan:2017}%
  \BibitemOpen
  \bibfield  {author} {\bibinfo {author} {\bibfnamefont {T.}~\bibnamefont
  {Tahamtan}}\ and\ \bibinfo {author} {\bibfnamefont {O.}~\bibnamefont
  {Sv{\'i}tek}},\ }\href {https://doi.org/10.1140/epjc/s10052-017-4945-z}
  {\bibfield  {journal} {\bibinfo  {journal} {Eur. Phys. J. C}\ }\textbf
  {\bibinfo {volume} {77}},\ \bibinfo {pages} {384} (\bibinfo {year}
  {2017})}\BibitemShut {NoStop}%
\bibitem [{\citenamefont {Frolov}\ and\ \citenamefont
  {Fursaev}(2005)}]{Frolov:2005}%
  \BibitemOpen
  \bibfield  {author} {\bibinfo {author} {\bibfnamefont {V.~P.}\ \bibnamefont
  {Frolov}}\ and\ \bibinfo {author} {\bibfnamefont {D.~V.}\ \bibnamefont
  {Fursaev}},\ }\href {https://doi.org/10.1103/PhysRevD.71.104034} {\bibfield
  {journal} {\bibinfo  {journal} {Phys. Rev. D}\ }\textbf {\bibinfo {volume}
  {71}},\ \bibinfo {pages} {104034} (\bibinfo {year} {2005})}\BibitemShut
  {NoStop}%
\bibitem [{\citenamefont {Frolov}\ \emph {et~al.}(2005)\citenamefont {Frolov},
  \citenamefont {Israel},\ and\ \citenamefont {Zelnikov}}]{Frolov1:2005}%
  \BibitemOpen
  \bibfield  {author} {\bibinfo {author} {\bibfnamefont {V.~P.}\ \bibnamefont
  {Frolov}}, \bibinfo {author} {\bibfnamefont {W.}~\bibnamefont {Israel}},\
  and\ \bibinfo {author} {\bibfnamefont {A.}~\bibnamefont {Zelnikov}},\ }\href
  {https://doi.org/10.1103/PhysRevD.72.084031} {\bibfield  {journal} {\bibinfo
  {journal} {Phys. Rev. D}\ }\textbf {\bibinfo {volume} {72}},\ \bibinfo
  {pages} {084031} (\bibinfo {year} {2005})}\BibitemShut {NoStop}%
\bibitem [{\citenamefont {Podolsk\'y}\ and\ \citenamefont
  {Ortaggio}(2003)}]{Podolsky:2003}%
  \BibitemOpen
  \bibfield  {author} {\bibinfo {author} {\bibfnamefont {J.}~\bibnamefont
  {Podolsk\'y}}\ and\ \bibinfo {author} {\bibfnamefont {M.}~\bibnamefont
  {Ortaggio}},\ }\href {https://doi.org/10.1088/0264-9381/20/9/307} {\bibfield
  {journal} {\bibinfo  {journal} {Classical and Quantum Gravity}\ }\textbf
  {\bibinfo {volume} {20}},\ \bibinfo {pages} {1685} (\bibinfo {year}
  {2003})}\BibitemShut {NoStop}%
\bibitem [{\citenamefont {Kadlecov\'a}\ \emph {et~al.}(2009)\citenamefont
  {Kadlecov\'a}, \citenamefont {Zelnikov}, \citenamefont
  {Krtou\ifmmode~\check{s}\else \v{s}\fi{}},\ and\ \citenamefont
  {Podolsk\'y}}]{Kadlecova:2009}%
  \BibitemOpen
  \bibfield  {author} {\bibinfo {author} {\bibfnamefont {H.}~\bibnamefont
  {Kadlecov\'a}}, \bibinfo {author} {\bibfnamefont {A.}~\bibnamefont
  {Zelnikov}}, \bibinfo {author} {\bibfnamefont {P.}~\bibnamefont
  {Krtou\ifmmode~\check{s}\else \v{s}\fi{}}},\ and\ \bibinfo {author}
  {\bibfnamefont {J.}~\bibnamefont {Podolsk\'y}},\ }\href
  {https://doi.org/10.1103/PhysRevD.80.024004} {\bibfield  {journal} {\bibinfo
  {journal} {Phys. Rev. D}\ }\textbf {\bibinfo {volume} {80}},\ \bibinfo
  {pages} {024004} (\bibinfo {year} {2009})}\BibitemShut {NoStop}%
\bibitem [{\citenamefont {Chakraborty}\ and\ \citenamefont
  {Kar}(2020)}]{Chak:2020}%
  \BibitemOpen
  \bibfield  {author} {\bibinfo {author} {\bibfnamefont {I.}~\bibnamefont
  {Chakraborty}}\ and\ \bibinfo {author} {\bibfnamefont {S.}~\bibnamefont
  {Kar}},\ }\href {https://doi.org/10.1103/PhysRevD.101.064022} {\bibfield
  {journal} {\bibinfo  {journal} {Phys. Rev. D}\ }\textbf {\bibinfo {volume}
  {101}},\ \bibinfo {pages} {064022} (\bibinfo {year} {2020})}\BibitemShut
  {NoStop}%
\bibitem [{\citenamefont {Bini}\ \emph
  {et~al.}(2018{\natexlab{a}})\citenamefont {Bini}, \citenamefont {Chicone},\
  and\ \citenamefont {Mashhoon}}]{Bini:2018}%
  \BibitemOpen
  \bibfield  {author} {\bibinfo {author} {\bibfnamefont {D.}~\bibnamefont
  {Bini}}, \bibinfo {author} {\bibfnamefont {C.}~\bibnamefont {Chicone}},\ and\
  \bibinfo {author} {\bibfnamefont {B.}~\bibnamefont {Mashhoon}},\ }\href
  {https://doi.org/10.1103/PhysRevD.97.064022} {\bibfield  {journal} {\bibinfo
  {journal} {Phys. Rev. D}\ }\textbf {\bibinfo {volume} {97}},\ \bibinfo
  {pages} {064022} (\bibinfo {year} {2018}{\natexlab{a}})}\BibitemShut
  {NoStop}%
\bibitem [{\citenamefont {Firouzjahi}\ and\ \citenamefont
  {Mashhoon}(2019)}]{Mashhoon:2019}%
  \BibitemOpen
  \bibfield  {author} {\bibinfo {author} {\bibfnamefont {H.}~\bibnamefont
  {Firouzjahi}}\ and\ \bibinfo {author} {\bibfnamefont {B.}~\bibnamefont
  {Mashhoon}},\ }\href {https://doi.org/10.1103/PhysRevD.100.084027} {\bibfield
   {journal} {\bibinfo  {journal} {Phys. Rev. D}\ }\textbf {\bibinfo {volume}
  {100}},\ \bibinfo {pages} {084027} (\bibinfo {year} {2019})}\BibitemShut
  {NoStop}%
\bibitem [{\citenamefont {Bini}\ \emph
  {et~al.}(2018{\natexlab{b}})\citenamefont {Bini}, \citenamefont {Chicone},
  \citenamefont {Mashhoon},\ and\ \citenamefont {Rosquist}}]{Bini_spin:2018}%
  \BibitemOpen
  \bibfield  {author} {\bibinfo {author} {\bibfnamefont {D.}~\bibnamefont
  {Bini}}, \bibinfo {author} {\bibfnamefont {C.}~\bibnamefont {Chicone}},
  \bibinfo {author} {\bibfnamefont {B.}~\bibnamefont {Mashhoon}},\ and\
  \bibinfo {author} {\bibfnamefont {K.}~\bibnamefont {Rosquist}},\ }\href
  {https://doi.org/10.1103/PhysRevD.98.024043} {\bibfield  {journal} {\bibinfo
  {journal} {Phys. Rev. D}\ }\textbf {\bibinfo {volume} {98}},\ \bibinfo
  {pages} {024043} (\bibinfo {year} {2018}{\natexlab{b}})}\BibitemShut
  {NoStop}%
\bibitem [{\citenamefont {Griffiths}\ and\ \citenamefont
  {Podolsky}(2009)}]{Griffiths:2009}%
  \BibitemOpen
  \bibfield  {author} {\bibinfo {author} {\bibfnamefont {J.~B.}\ \bibnamefont
  {Griffiths}}\ and\ \bibinfo {author} {\bibfnamefont {J.}~\bibnamefont
  {Podolsky}},\ }\href {https://doi.org/10.1017/CBO9780511635397} {\emph
  {\bibinfo {title} {{Exact Space-Times in Einstein's General Relativity}}}}\
  (\bibinfo  {publisher} {Cambridge Univ. Press},\ \bibinfo {address}
  {Cambridge, England},\ \bibinfo {year} {2009})\BibitemShut {NoStop}%
\bibitem [{\citenamefont {O'Loughlin}\ and\ \citenamefont
  {Demirchian}(2019)}]{Loughlin:2019}%
  \BibitemOpen
  \bibfield  {author} {\bibinfo {author} {\bibfnamefont {M.}~\bibnamefont
  {O'Loughlin}}\ and\ \bibinfo {author} {\bibfnamefont {H.}~\bibnamefont
  {Demirchian}},\ }\href {https://doi.org/10.1103/PhysRevD.99.024031}
  {\bibfield  {journal} {\bibinfo  {journal} {Phys. Rev. D}\ }\textbf {\bibinfo
  {volume} {99}},\ \bibinfo {pages} {024031} (\bibinfo {year}
  {2019})}\BibitemShut {NoStop}%
\bibitem [{\citenamefont {Bhattacharjee}\ \emph {et~al.}(2019)\citenamefont
  {Bhattacharjee}, \citenamefont {Kumar},\ and\ \citenamefont
  {Bhattacharyya}}]{Srijit:2019}%
  \BibitemOpen
  \bibfield  {author} {\bibinfo {author} {\bibfnamefont {S.}~\bibnamefont
  {Bhattacharjee}}, \bibinfo {author} {\bibfnamefont {S.}~\bibnamefont
  {Kumar}},\ and\ \bibinfo {author} {\bibfnamefont {A.}~\bibnamefont
  {Bhattacharyya}},\ }\href {https://doi.org/10.1103/PhysRevD.100.084010}
  {\bibfield  {journal} {\bibinfo  {journal} {Phys. Rev. D}\ }\textbf {\bibinfo
  {volume} {100}},\ \bibinfo {pages} {084010} (\bibinfo {year}
  {2019})}\BibitemShut {NoStop}%
\end{thebibliography}%

\end{document}